\documentclass[preprint,amsmath,amssymb,aps,prd,showpacs]{revtex4-1}
 \pdfoutput=1

\usepackage{float}

\usepackage{epsfig}
\usepackage{dcolumn}% Align table columns on decimal point
\usepackage{bm}% bold math
\usepackage{tikz}
\usepackage{graphicx, graphics, color}
\usepackage{dcolumn}% Align table columns on decimal point
\usepackage{bm}% bold math
\usepackage{hyperref}% add hypertext capabilities
\usepackage[mathlines]{lineno}
\usepackage{float}
\usepackage{subfig}

\newcommand{\be}{\begin{equation}}
\newcommand{\ee}{\end{equation}}
\newcommand{\ben}{\begin{eqnarray}}
\newcommand{\een}{\end{eqnarray}}

\newcommand{\cL}{{\cal L}}

\newcommand{\cE}{{\cal E}}

\newcommand{\na}{\nabla}

\newcommand{\tA}{\tilde A}
\newcommand{\tF}{\tilde F}

\newcommand{\ep}{\epsilon}

\newcommand{\ga}{\gamma}

\newcommand{\tB}{{\tilde B}}

\newcommand{\te}{\tilde e}

\graphicspath{{figs/}}
\graphicspath{{../figs/}}

\begin{document} 

\title{ Can synchrotron radiation reveal the presence of {\it dark sector} around black hole?}

%\author{Bartlomiej Kiczek} 
%\email{bkiczek@kft.umcs.lublin.pl}
\author{Marek Rogatko} 
\email{rogat@kft.umcs.lublin.pl}
\author{Paritosh Verma} 
\email{pverma@kft.umcs.lublin.pl}
%\author{Karol I. Wysokinski}
%\email{karol@tytan.umcs.lublin.pl}
\affiliation{Institute of Physics, % \protect \\
Maria Curie-Sklodowska University, % \protect \\
pl.~Marii Curie-Sklodowskiej 1,  20-031 Lublin,  Poland}

%\author{Bartlomiej Kiczek} 
%\email{bkiczek@kft.umcs.lublin.pl}
%\author{Marek Rogatko} 
%\email{rogat@kft.umcs.lublin.pl}
%,marek.rogatko@poczta.umcs.lublin.pl }
%\affiliation{Institute of Physics, % \protect \\
%Maria Curie-Sklodowska University, % \protect \\
%pl.~Marii Curie-Sklodowskiej 1,  20-031 Lublin,  Poland}

\date{\today}% It is always \today, today,
             %  but any date may be explicitly specified

%%%%%%%%%%%%%%%%%%%%%%%
\begin{abstract}
We studied synchrotron radiation of a massive charged under visible and hidden sector groups, moving in equatorial plane around spherically symmetric
weakly magnetized black hole. As a model of dark matter we choose the one, in which Maxwell field is coupled to the additional $U(1)$-gauge field envisaging
the dark sector, the so-called dark photon model. Magnetization of a black hole also stems from Maxwell-dark photon electrodynamics. One found the radiation power
and energy loss of the particle and looked for the imprints of dark matter on those phenomena.
\end{abstract}
%%%%%%%%%%%%%%%%%%%%%%%

\maketitle
\flushbottom

%%%%%%%%%%%%%%%%%%%%%%%%%%%%%%%%%%%%%%%%%%%%%%%%%%%%%%%%%%%%%%%%%%%%%%%%%%%
%%%%%%%%%%%%%%%%%%%%%%%%%%%%%%%%%%%%%%%%%%%%%%%%%%%%%%%%%%%%%%%%%%%%%%%%%%%
\section{Introduction}
\label{sec:intro}
An ultra relativistic charged particle moving in a magnetic field experiences the Lorentz force and emits electromagnetic radiation due to time varying dipole moment. If the velocity is orthogonal to acceleration, one observes synchrotron radiation. 
This phenomenon plays a vital role in particle accelerators. Moreover,
motion of a test particle in various black hole spacetimes, has been widely elaborated.
For instance, synchrotron radiation from a charged particle travelling
in a circular nongeodesic orbit around a weakly magnetized Schwarzschild and Kerr black hole, were scrutinized in \cite{Khriplovich1974, ali81, Sokolov.et.al}. 

High-frequency electromagnetic and gravitational radiation from a relativistic particles falling into a Schwarzschild and a Kerr black hole \cite{Dymnikova1977}, electromagnetic radiation from a radially free falling monopole and point-like dipole 
into a Schwarzschild black hole \cite{Shatskiy2013}, the radiation from free falling dipole  \cite{Shatskiy.et.al}, the bremsstrahlung of a dipole falling into Schwarzschild black hole  \cite{sha15}, as well as,
an electromagnetic radiation spectrum from charged particles moving gently in an eccentric bound equatorial orbit around a static black hole immersed into a weak dipolar magnetic field \cite{Papad2015}, were under inspections.

Additionally, in \cite{sho15}, synchrotron radiation from a charged particle moving in an equatorial palne around Schwarzschild black hole, under the assumption that the dynamic of the particle was mainly 
driven by electromagnetic forces resulting in nongeodesic particle world line, was elaborated.

On the other hand, synchrotron radiation in odd dimensional spacetimes has been studied in \cite{gal20}, while the geometric optics approximation has been implemented in
studies of polarization direction and luminosity of synchrotron radiation of moving electron in curved spacetime \cite{hu22}.

%%%%%%%%%%%%%%%%%%%%%%%%%%%%%%
Physics of {\it dark matter} and {\it dark energy} is still one of the prominent puzzles of modern physics \cite{planck}. 
Despite various ambiguous indications of {\it dark sectors}, neither we have revealed them experimentally nor we have a specific theory to elucidate them. The Standard Model is also 
unsuccessful in expounding the {\it dark sector} theoretically.
However, one of the rudimentary solutions is to assume the presence of particles that interact weakly with the ordinary matter in the Universe.

In this article, we consider the concept of {\it dark photons}, which is a persuasive competitor for explaining physics beyond the Standard Model. {\it Dark photon} is a 
$U(1)$ Abelian gauge boson which couples to theordinary Maxwell gauge field via {\it kinetic mixing term} \cite{hol86,cap21}. The idea in question has also its justification in 
the contemporary unification scheme  \cite{ach16}, where the mixing portals coupling the ordinary Maxwell 
and auxiliary gauge fields charged under their own groups,  are intensively elaborated. 

On the other hand, it has been claimed that {\it dark photons} might be responsible for several astrophysical observations like, e.g.,
anomalous astrophysical effects like $511~ keV$
gamma rays \cite{jea03}, excess of the positron cosmic ray flux in galaxies \cite{cha08},  the observations of an anomalous monochromatic 
$3.56~ keV$ X-ray line in the spectrum of some galaxy clusters \cite{bub14}. 
 
 One can observe ongoing search using astrophysical and laboratory experiments \cite{fil20}, 
to find the range of values for {\it dark photon} - Maxwell field coupling constant and the mass of the {\it hidden photon}. Namely,
studies of gamma rays emissions from dwarf galaxies \cite{ger15},
inspection of dilaton-like coupling to photons caused by ultra-light {\it dark matter} \cite{bod15} 
 fine structure constant oscillations \cite{til15}, {\it dark photon} emission during supernova event \cite{cha17}, 
 electron excitation measurements in CCD-like detector \cite{sensei}, the search for a {\it dark photon} in $e^+ e^-$ collisions at BABAR experiment \cite{lee14},
 measurements of the muon anomalous effect \cite{dav11}, are under inspection.

%%%%%%%%%%%%%%%%%%%%%%%%%%%%%%%%%%%%%%%%%%%%%%%%
Recent studies reveal \cite{rom23} -\cite{fil23}
that the new exclusion limit for the $\alpha$-coupling constant is given by $\alpha =1.6 \times
10^{-9}$ and the mass range of {\it dark photon}
$2.1\times 10^{-7} - 5.7 \times10^{-6} eV$ \cite{rom23} -\cite{fil23}. Moreover,
quantum limited amplification scrutinizes the kinetic mixing coupling constant to $10^{-12}$ level for majority of {\it dark photon} masses  \cite{ram23}. 
On the other hand, in Ref. \cite{kot23}, 
an upper bound on coupling constant  $\alpha < 0.3-2 \times 10^{-10}$ (at 95 percent confidence level)  has been established.

It turns out that {\it dark photon} may contribute to a new spectral anisotropies in the cosmic microwave background (CMB), the so-called patchy dark screening
\cite{piv24}. It is caused by the conversion of photons to the dark ones within large-scale structures, which takes place through the {\it kinetic mixing term}.
Analysis of Planck CMB and unWISE galaxy survey \cite{mcc24}
reveal constraints on {\it dark photon} parameter $\alpha \le 4.5 \times 10^{-8}$ (95 percent of confidence level),
as well as, the range of its mass $ 10^{-13}  eV \le m_{dark ~ph} \le 10^{-11} eV$.

Moreover, the absence of  evidence of the most popular {\it dark matter} candidates in the experimental data, triggers the diversification of the searching efforts
for {\it dark sector} \cite{ber18b}. Namely, perhaps one ought to take
 into account astronomical surveys, gravitational wave observatories, and new concepts of Earth experiments.
It happens that 
 one of the most promising way of finding the presence of {\it dark matter} may be studies of the environments of compact objects like black holes,
 wormholes and compact star-like objects. Some work in this direction, 
 connecting with {\it dark matter} clouds and their influence on compact objects were performed in 
 \cite{kic19}-\cite{kic22}. On the other hand, the uniqueness of black hole in {\it dark photon} Einstein-Maxwell gravity was treated in \cite{rog23,rog24}.

 %%%%%%%%%%%%%%%%%%%%%%%%%%%%%%%%%
The main aim of our study is to find potentially observable traces of the {\it dark matter} influence on synchrotron radiation of charged 
particle moving around weakly magnetized Schwarzschild black hole and to reveal effects which can distinguish the case under consideration in comparison to the ordinary Maxwell 
one.
Magnetization of the black hole in question will be connected with the ordinary Maxwell field and {\it dark photon} one.

%%%%%%%%%%%%%%%%%%%%%%%%%%%%%%%%%

The organization of the paper is as follows. In Sec. II we describe the main features of the {\it dark photon} theory in which the {\it hidden sector} is modelled by the auxiliary $U(1)$-gauge field coupled to the ordinary Maxwell one.
In Sec. III e present some basic features of the synchrotron radiation in the context of {\it dark photon} theory, finding the intensity of radiation and comparing it with the {\it visible
sector} one.  Sec. IV will be devoted to the studies of circular motion of the particle in the equatorial plane. By means of the Killing vectors admitted by the spacetime in question,
one defines conserved quantities and look for the influence of {\it dark matter} on the energy loss.
In Sec. V we conclude our investigations.

In our paper we use geometrized units $G=1,~c=1$, the line element has the signature $(- +++)$. In all the figures mass of black hole and particle, charge, and magnetic fields
are taken as dimensionless.

%%%%%%%%%%%%%%%%%%%%%%%%%%%%%%%%%%%%%%%%%%%%%%%%%%%%%%%%%%%%%%%%%%
\section{Model of dark sector} \label{sec:darkphoton}
This section will be devoted to the short description of {\it dark sector} model, namely in our studies we shall implement the {\it dark photon}
model in which the additional $U(1)$-gauge field represents the {\it hidden sector}. This auxiliary field interacts with Maxwell one via so-called {\it kinetic mixing} term 
being proportional to both gauge field strengths. The coupling constant between those two sectors will be denoted by $\alpha$. 
The action describing two coupled, massless gauge field is given by
\be
S_{M-dark~ photon} = \int  d^4x  \Big(
- F_{\mu \nu} F^{\mu \nu} - B_{\mu \nu} B^{\mu \nu} - {\alpha}F_{\mu \nu} B^{\mu \nu}
\Big).
\label{ac dm}
\ee  
In order to get rid ot the {\it kinetic mixing } term, which simplifies calculation to great extent), we define
\ben \label{transA}
\tA_\mu &=& \frac{\sqrt{2 -\alpha}}{2} \Big( A_\mu - B_\mu \Big),\\ \label{transB}
\tB_\mu &=& \frac{\sqrt{2 + \alpha}}{2} \Big( A_\mu + B_\mu \Big),
\een
which results in the modified action
\be
S_{M-dark~ photon}  = \int d^4x  \Big(
- \tF_{\mu \nu} \tF^{\mu \nu} - \tB_{\mu \nu} \tB^{\mu \nu}
\Big),
\label{vdc}
\ee
where we define $\tF_{\mu \nu} = 2 \na_{[\mu }\tA_{\nu ]}$, and respectively $\tB_{\mu \nu} = 2 \na_{[\mu }\tB_{\nu ]}$. Equating functional derivatives of
of the action (\ref{vdc}) with respect to $\tA_\mu$ and $\tB_\mu$ exhibits the equations of motion for Maxwell-{\it dark matter} system given by
\be
\na_{\mu} \tF^{\mu \nu } = 0, \qquad \na_{\mu} \tB^{\mu \nu } = 0.
\label{fb}
\ee
Further, due to the consistency of the approach one redefines the charges coupled to the $A_\mu$ and $B_\mu$ gauge fields in the manner
as follows:
\ben \label{cA}
\te_A &=& \frac{\sqrt{2 -\alpha}}{2} \Big( e - e_d \Big),\\ \label{cB}
\te_B &=& \frac{\sqrt{2 + \alpha}}{2} \Big( e + e_d \Big),
\een
where $e$ denotes charge bounded with ordinary Maxwell field, while $e_d$ charge connected with {\it dark sector}. In what follows we assume that
the charges are positive in order to gain comparison with the synchrotron radiation in the ordinary Maxwell case studied in Ref. \cite{sho15}.

he equation of motion of charged massive particle in the background of the modified gauge fields results from the action provided by
\be
S = - \int m \sqrt{- ds^2} + \te_A \int \tA_{\mu} dx^\mu + \te_B \int \tB_\mu dx^\mu,
\ee
and yields the dynamical equation of the form
\be
m~\frac{D u^\mu}{d \tau} = \Big( \te_A \tF^{\mu \nu} +  \te_B \tB^{\mu \nu} \Big) u_\nu,
\label{eqmot}
\ee
where $\frac{D}{d \tau}$ stands in general case for the covariant derivative with respect to the proper time, while $u_\mu$ stands for the particle four velocity,
wthe normalization condition $u_\alpha u^\alpha = -1$.

On the other hand, the four-momentum of the massive particle is given by
\be
{\tilde p}_\mu = m u_\mu +\te_A \tA_\mu +  \te_B  \tB_\mu.
\label{mom}
\ee
%%%%%%%%
Having in mind the action (\ref{ac dm}), 
the above equation provides the consistency in choosing transformed charges (\ref{cA})-(\ref{cB}) and fields (\ref{transA}) and (\ref{transB}).

It happens that in the case, when the dark charge is equal to zero, the above relation reduces to the following:
\be
{\tilde p}_\mu = m u_\mu +e \Big(A_\mu + \frac{\alpha}{2}B_\mu \Big).
\ee
From the above relation it can be seen the modifications of the kinetic momentum introduced by {\it dark sector} gauge field.

%%%%%%%%%%%%%%%%%%%%%%%%%%%%%%%%%%%%%%%%%%%%%%%%%%%%%%%%%%%%%%%

%%%%%%%%%%%%%%%%%%%%%%%%%%%%%%%%%%%%%%%%%%%%%%%%%%%%%%%%%%%%%%%%%%%
\subsection{Weakly magnetized black holes in dark photon theory}
As was mentioned in the introduction, our main goal was to study synchrotron radiation in the spacetime of
spherically symmetric black hole. 
In this subsection, for the readers' convenience we recall the basic points in the derivation of weakly magnetized gauge components in {\it dark photon} theory.

In order to find the values of both gauge fields in our theory, we shall follow the method presented
in Ref. \cite{wal74}, where it has been shown that
the gauge potential can be defined as a linear combination of the adequate Killing vector field responsible for the symmetries in the spacetime. Next due to the fact that
both Killing vector, in the case of vanishing Ricci tensor $R_{\mu \nu}$, and gauge potentials in Lorenz gauge satisfy the similar equations.

In the considered Einstein-Maxwell {\it dark photon} theory we have Killing vector fields $\xi^\mu_{(t)}$ and $\xi^\mu_{(\phi)}$ which
generate time translation and rotations around the symmetry axis, and Lorenz gauge enables us to write that $\na_\beta \na ^\beta \tA_\mu = \na_\beta \na ^\beta \tB_\mu =0$.
Moreover one has that
\ben
\tA^\mu &=& \ga_1~ \xi^\mu_{(t)} + \ga_2 ~\xi^\mu_{(\phi)},\\ 
\tB^\mu &=&  \delta_1~ \xi^\mu_{(t)} + \delta_2 ~\xi^\mu_{(\phi)}.
\een
The definitions of ADM mass, angular momentum and charges
\be
- 8 \pi M = \int \ep_{\alpha \beta \ga \delta} \na^\ga \xi^\delta_{(t)}, 
\ee
 angular momentum
 \be
16 \pi J = \int \ep_{\alpha \beta \ga \delta} \na^\ga \xi^\delta_{(\phi)},
\ee
and charges bounded with two sectors
\be
4 \pi Q(\tF) = \int \ep_{\alpha \beta \ga \delta} \tF^{\ga \delta}, \qquad 
4 \pi Q(\tB) = \int \ep_{\alpha \beta \ga \delta} \tB^{\ga \delta},
\label{charges}
\ee
enables one to find the relations for gauge potentials
\ben
\tA^\mu &=& \frac{B^{(\tF)}_0}{2} \Big[ \xi^\mu_{(\phi)} + 2~a \xi^\mu_{(t)} \Big] - \frac{Q(\tF)}{2 M} \xi^\mu_{(t)},\\
\tB^\mu &=& \frac{B^{(\tB)}_0}{2} \Big[ \xi^\mu_{(\phi)} + 2~a \xi^\mu_{(t)}  \Big] - \frac{Q(\tB)}{2 M} \xi^\mu_{(t)},
\een
where $a = J/M$. For the consistency with the later considerations, one has that the mixture of magnetic 
fields components connected with {\it visible} and {\it dark sectors} is written as follows:
\be
B^{(\tF)}_0 = \frac{\sqrt{2 -\alpha}}{2} \Big( B^{(F)}_0 - B^{(B)}_0 \Big),
\qquad
B^{(\tB)}_0 = \frac{\sqrt{2 + \alpha}}{2} \Big( B^{(F)}_0 + B^{(B)}_0 \Big).
\label{bbb}
\ee
Consequently from the definition of the charges (\ref{charges}) we arrive at
\be
Q{(\tF)} = \frac{\sqrt{2 -\alpha}}{2} \Big( Q{(F)} - Q{(B)} \Big),
\qquad
Q{(\tB)} = \frac{\sqrt{2 + \alpha}}{2} \Big( Q{(F)} + Q{(B)} \Big),
\ee
where the charges $Q(F)$ and $Q(B)$ are defined as those in the equation (\ref{charges}) but now for Maxwell and {\it dark photon}, respectively.

%%%%%%%%%%%%%%%%%%%%%%%
For the case when $a = J/M =0$, one obtains the static black hole case. Moreover if $Q=0$
 the mixture of gauge field components is given by
 \be 
 \tA^\mu = \frac{ B^{(\tF)}_0}{2} \xi^\mu_{(\phi)}, \qquad  \tB^\mu = \frac{B^{(\tB)}_0}{2} \xi^\mu_{(\phi)},
\label{stat}
\ee
which implies that
$A^\mu = \frac{ B^{(F)}_0}{2} \xi^\mu_{(\phi)}$ and $B^\mu = \frac{B^{(B)}_0}{2} \xi^\mu_{(\phi)}$.

%%%%%%%%%%%%%%%%%%%%%%%%%%%%%%%%%%%%%%%%%%%%%%%%%%%%%%%%%%
\section{Synchrotron radiation with dark sector}
In order to derive the relation for the total radiation coming from the motion of charged particle (charged under Maxwell and {\it dark photon} $U(1)$ groups)
in magnetic fields stemming from both two sectors, let us discuss first the non-relativistic case. For the case of Maxwell and {\it dark photon} fields,
the total radiation can be found by considering $\ddot{\vec d}_{(\tF)}= \te_A \ddot{\vec r}$,~ $\ddot{\vec d}_{(\tB)} = \te_B \ddot{\vec r}$ and integrating the expression
\be
dI = \frac{1}{4 \pi} \Big[ \Big( \ddot{\vec d}_{(\tF)} \times {\vec n} \Big)^2 + \Big( \ddot{\vec d}_{(\tB)} \times {\vec n} \Big)^2 \Big] d \Omega,
\ee
over $\theta$ , one gets
\be
I = I_{(\tF)} + I_{(\tB)} = \frac{2}{3} \frac{d^2 \vec r}{dt^2} \Big(  \te_A^2 + \te_B^2 \Big).
\label{nonrelrad}
\ee
The above formula (\ref{nonrelrad}) is derived under the assumption that the velocity of the particle is very small comparing to the light velocity and is not 
applicable to the relativistic description. However one can consider the particle which is at rest in the reference frame, at given time. In this particular frame
the particle radiates in the time $dt$ the energy 
\be
dE = \frac{2}{3}  \frac{d^2 \vec r}{dt^2} \Big(  \te_A^2 + \te_B^2 \Big) dt.
\ee
In order to transform this expression to the arbitrary reference frame, one rewrites it in four-dimensional form. Now the radiated four-momentum yields
\be
dP^\alpha =  \frac{2}{3}  \Big(  \te_A^2 + \te_B^2 \Big) 
\frac{d u^\beta}{d \tau} \frac{d u_\beta}{d \tau} dx^\alpha, 
\ee
where $u^\beta$ denotes four-velocity. Integrating the above equation we arrive at the total four-momentum radiated during the passage of time by a particle charge under two sectors,
moving due to its equation of motion connected with the terms proportional to the square of $du^\beta/d \tau$.

However the application to the curved spacetime problem meets some obstacles.
One has to recall that because of the fact that radiation process is of a non-local nature, the direct application of the Minkowski-like formula
derived for radiated four-momentum by a charge particle, is not suitable for curved spacetime, in a general case.
However if the non-gravitational forces play the crucial role, the studied particle world-line is not a geodesic \cite{ali81}-\cite{sho15}. This fact enables us to investigate the
emergence of radiation (electromagnetic and {\it dark photon} ones) as a local phenomena, described by local quantities. This conclusion will be the key point in our derivations of 
synchrotron radiation modified by {\it dark matter} sector.

Thus, the radiated four-momentum of a particle charged under ordinary (Maxwell) and {\it dark sector} groups, moving around a weakly magnetized Schwarzschild black hole yields
\be
dP^\beta =
\frac{2}{3} \Big( \te_A^2 + \te_B^2 \Big) \frac{D u^\mu}{d \tau}\frac{D u_\mu}{d \tau} dx^\beta,
\ee
where the equation of motion $\frac{D u^\mu}{d \tau}$ are given by (\ref{eqmot}). $dP^\beta$ is measured by a static observer situated at flat asymptotic region

On the other hand, due to the symmetries of the spherical static spacetime, we have as has been mentioned, two Killing vectors $\xi^\mu_{(t)}$ and $\xi^\mu_{(\phi)}$.
The gauge fields should be invariant due to the symmetries generated by them, i.e.,
\ben
\cL_{\xi^\mu_{(t)}} \tA_\ga &=& 0, \qquad \cL_{\xi^\mu_{(\phi)}} \tA_\ga = 0, \\
\cL_{\xi^\mu_{(t)}} \tB_\ga &=& 0, \qquad \cL_{\xi^\mu_{(\phi)}} \tB_\ga = 0.
\een
The same relations ought to be fulfilled by $A_\beta$ and $B_\beta$ gauge fields. It implies that the corresponding strength tensors
are provide by
\ben
F_{\mu \nu} &=& 2  B_0^{(F)} r \sin \theta \Big( \sin \theta~ \delta_{[ \mu}^r \delta_{\nu ]}^\phi + r \cos \theta ~ \delta_{[ \mu}^\theta \delta_{\nu ]}^\phi \Big), \\
 B_{\mu \nu} &=& 2  B_0^{(B)} r \sin \theta \Big( \sin \theta~ \delta_{[ \mu}^r \delta_{\nu ]}^\phi + r \cos \theta~  \delta_{[ \mu}^\theta \delta_{\nu ]}^\phi \Big),
 \een

Finally in the equatorial plane $\theta = \pi/2$, for the ultra-relativistic approximation, the radiated four-momentum influenced by {\it dark sector} gauge field, is provided by
\be
dP^\beta =
\frac{2 e^4}{3 m^2} \Big( 1 - \frac{r_g}{r} \Big)^2 \Big( \frac{dt}{d\tau} \Big)^2 \Big[ B_0^{(F)} + \frac{\alpha}{2} B_0^{(B)} \Big]^2 dx^\beta.
\ee

Further, calculating the spectral distribution function like in Refs. \cite{ali81}-\cite{sho15}, for the synchrotron radiation effected by {\it dark sector} field, one 
arrives at the following relation:
\be
\frac{d I}{dz} = \frac{3 \sqrt{3}}{4 \pi} \frac{e^4}{m^2} \Big( \frac{dt}{d\tau} \Big)^2 \Big[ B_0^{(F)} + \frac{\alpha}{2} B_0^{(B)} \Big]^2 F(z),
\label{spec}
\ee
where $F(z) = z \int_z^\infty K_{5/3}(y)dy$, while $z = \omega/\omega_c$ is equal to
\be
\omega_c = \frac{3 e \Big( B_0^{(F)} + \frac{\alpha}{2} B_0^{(B)} \Big)}{2 m} r^2 ~\Big( 1 - \frac{r_g}{r} \Big)^2.
\ee

Due to the properties of the MacDonald function $K_{5/3}(z)$, the maximum of the function $F(z)$ is located near $z \approx 0.29$,
the main part of the synchrotron radiation is given around
\be
\omega = 0.29~ \omega_c(Maxwell)~\Bigg[ 1 + \frac{\alpha}{2}~\frac{B_0^{(B)}}{B_0^{(F)}} \Bigg].
\ee
One can observe that the shift in $\omega$ is proportional to the ratio of magnetic {\it dark matter} field to magnetic Maxwell field.

In Fig. 1 one considers $\omega_c$ as a function of $r$-coordinate, with fix values of black hole and particle masses, with the dominant value of {\it dark photon} magnetic field.
One observes that the larger value of $\alpha$-coupling constant we take, the larger $\omega_c$ one has.

%%%%%%%%%%%%%%%%%%%%%%%%%%%%%%%%%%%
\begin{figure}[H]   
\centerline{ \includegraphics[width=12cm]{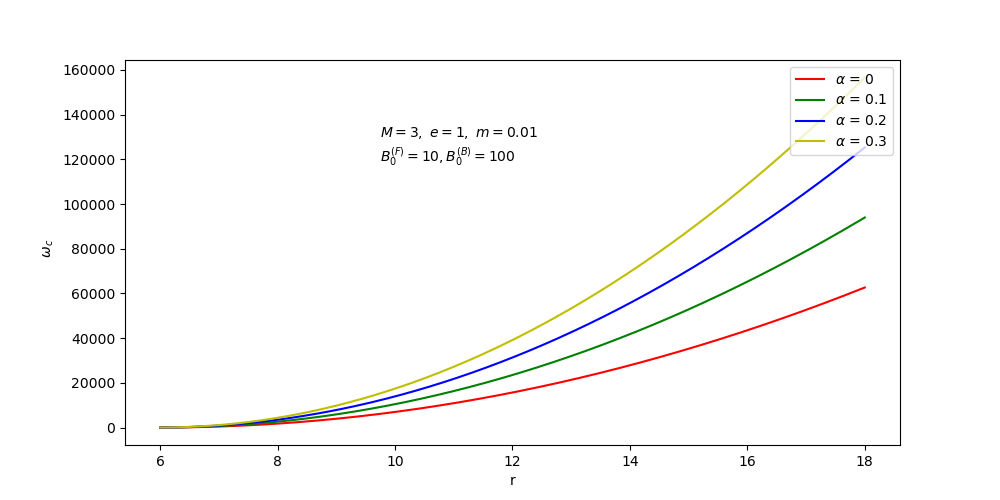} }
\caption{(color on line) The radial dependence of $\omega_c$ for various values of coupling constant, the case when { \it dark sector} fields is dominated.}
\label{Fig:1}
\end{figure}

The same behavior for the magnitude of $B_0^{(F)} = B_0^{(B)}$, but the distances among the curves responsible for different $\alpha's$ are smaller, may be observed in Fig. 2.

\begin{figure}[H]   
\centerline{ \includegraphics[width=12cm]{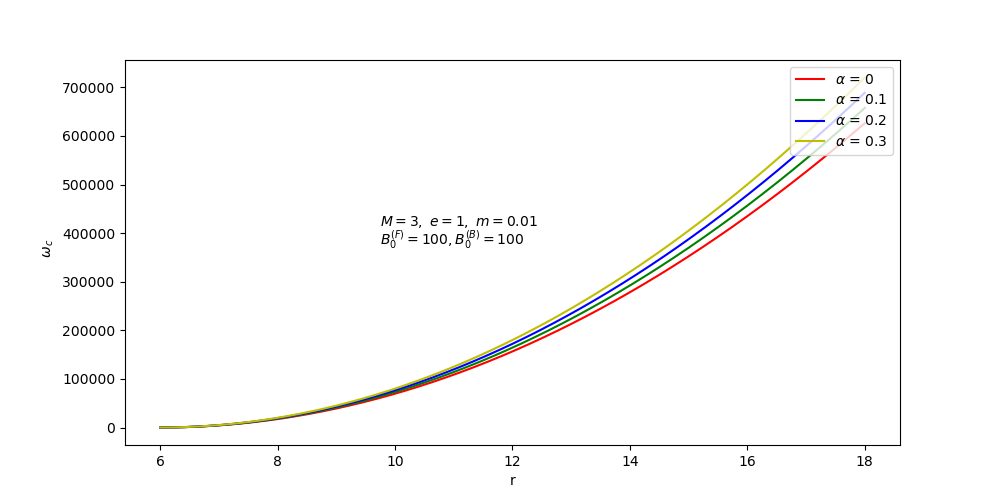} }
\caption{(color on line) The same dependence as in Fig.1, but magnetic fields of both sector have the same values.}
\label{Fig:2}
\end{figure}

However in the case when Maxwell magnetic field is greater that the {\it dark sector} one, one can hardly recognizes the curves for different coupling constant.
\begin{figure}[H]   
\centerline{ \includegraphics[width=12cm]{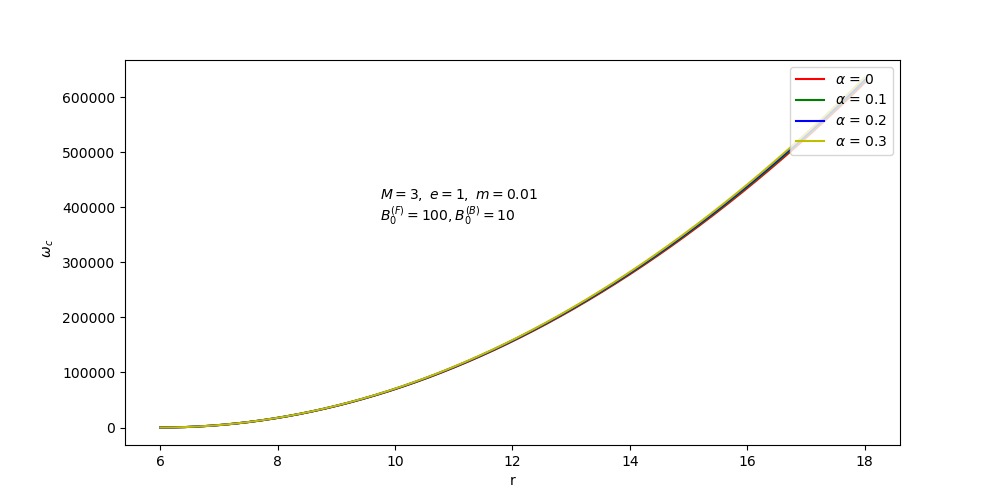} }
\caption{(color on line) The radial dependence of $\omega_c$, for different values of $\alpha$-coupling constant,  Maxwell fields dominates around black hole.}
\label{Fig:3}
\end{figure}

%%%%%%%%%%%%%%%%%%%%%%%%%%%%%%%%%%%%

On the other hand, the radiation intensity is given by the following expression, derived by the integration of the equation (\ref{spec}) over the whole range of
variable $z$, i.e., from zero to infinity.
It yields
\ben \label{rad}
I &=& \frac{2}{3} \frac{e^4}{m^2} \Big( \frac{dt}{d\tau} \Big)^2 \Big[ B_0^{(F)} + \frac{\alpha}{2} B_0^{(B)} \Big]^2 ~\Big( 1 - \frac{r_g}{r} \Big)^3 \\ \nonumber
&=& I(Maxwell) + I({dark ~matter}),
\een
where in $I({dark ~matter})$ we get term proportional to $\alpha$ and $\alpha^2$ coupling constant.

For the ultra-relativistic particle moving in the spacetime of static spherically symmetric weakly magnetized black hole influenced by {\it dark photon}, the total radiation intensity
is greater compared to the Maxwell case, i.e., the {\it dark matter sector} causes the increase of the synchrotron radiation. Of course, the effect is small (because of the
estimations of $\alpha$-coupling parameter value) but it indicates the presence of {\it dark matter} around the aforementioned black hole.

In the case of radial dependence of synchrotron radiation we have almost the same conclusions. Namely, Fig. 4, for the dominant values of Maxwell field we can not distinguish the curves, for $B_0^{(F)} = B_0^{(B)}$ (Fig. 5)
one has different curves with small distances among them, while for the dominant  case of {\it dark sector}, one obtains different curves with large distances between them (Fig. 6).

\begin{figure}[H]   
\centerline{ \includegraphics[width=12cm]{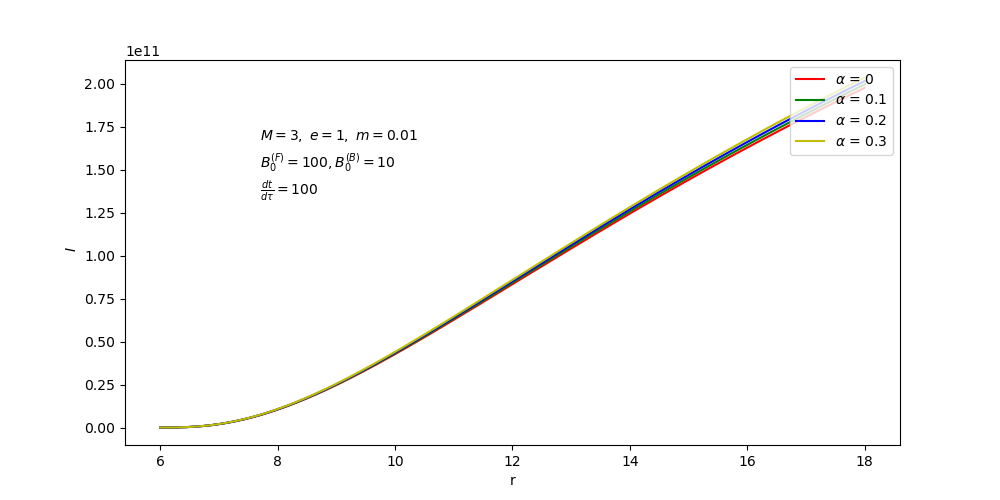} }
\caption{(color on line) Intensity of synchrotron radiation, as a function of the distance $r$, for different values of $\alpha$, when Maxwell field is greater comparing to {\it dark photon} one. }
\label{Fig:4}
\end{figure}

\begin{figure}[H]   
\centerline{ \includegraphics[width=12cm]{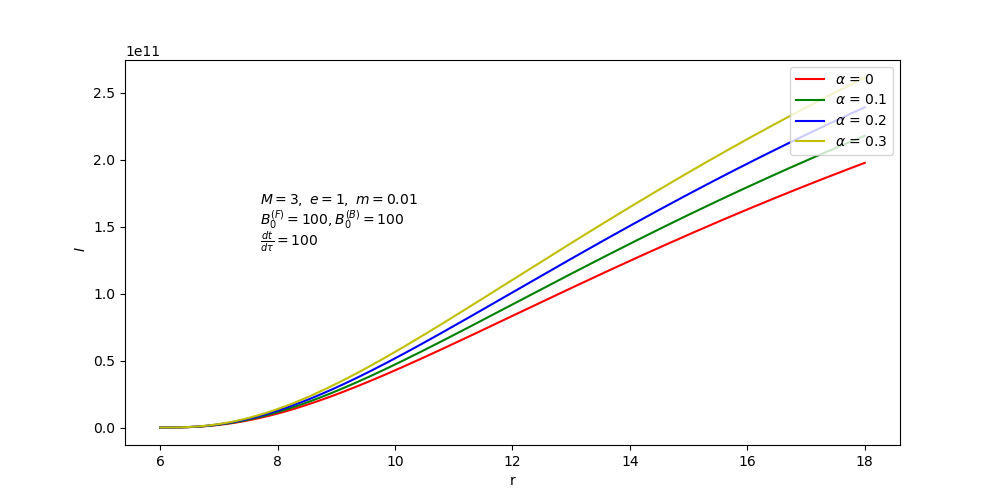} }
\caption{(color on line) Intensity of synchrotron radiation when both sector fields have the same magnitude. }
\label{Fig:5}
\end{figure}

\begin{figure}[H]   
\centerline{ \includegraphics[width=12cm]{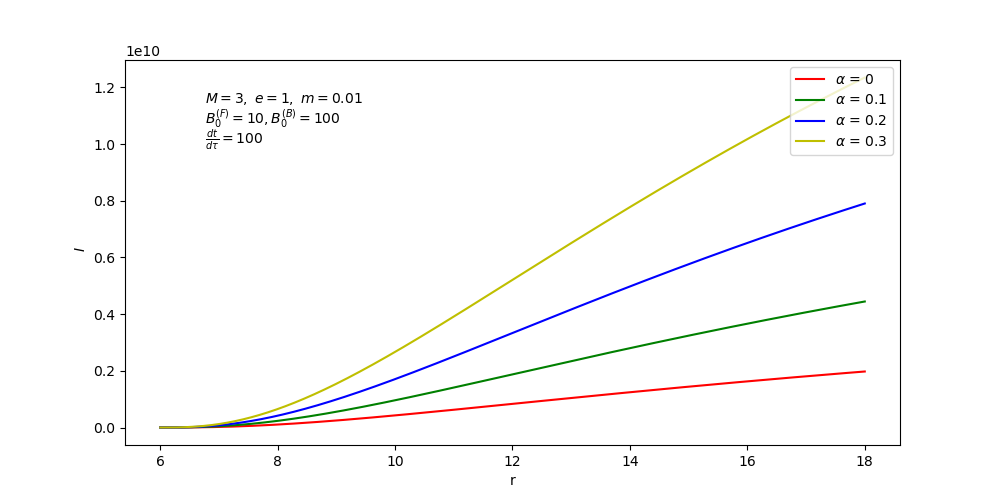} }
\caption{(color on line) Intensity of synchrotron radiation for the  ascendancy of {\it dark photon} field.}
\label{Fig:}
\end{figure}

For all those quantities, the bigger value of $\alpha$ we take, the bigger $\omega_c$ and intensity of radiation one achieves.

Let us fix the particle mass and charge and look for the influence of black hole mass. In Fig. 7, one has that the more massive black hole is the larger value of intensity we have.

\begin{figure}[H]   
\centerline{ \includegraphics[width=12cm]{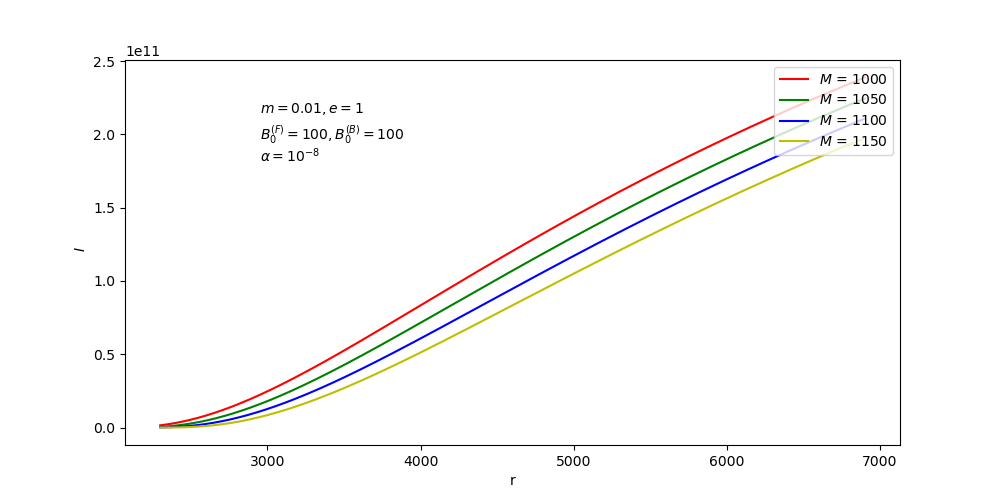} }
\caption{(color on line) Dependence of intensity on the distance, for different values of black hole mass. }
\label{Fig:15}
\end{figure}

On the other hand, Fig. 8, having black hole mass fixed and changing particle mass, one can draw a conclusion that the smaller mass of the particle we take the greater value of intensity of synchrotron radiation we arrive at. Moreover, the growth of $\alpha$ value
cause the amplification of intensity.

\begin{figure}[H]   
\centerline{ \includegraphics[width=12cm]{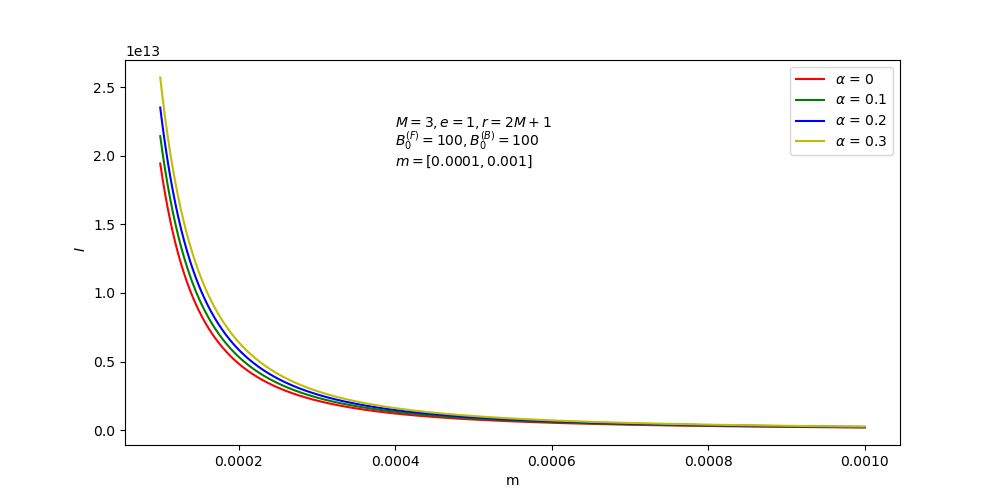} }
\caption{(color on line) Intensity as a function of the particle mass. }
\label{Fig:15}
\end{figure}

The radial dependence of synchrotron intensity is also plotted in Fig. 9, when we set black hole mass, particle charge and Maxwell field is greater than {\it dark photon} one.
The studies of different masses of particle reveal that the smaller mass we take the larger value we achieve.

\begin{figure}[H]   
\centerline{ \includegraphics[width=12cm]{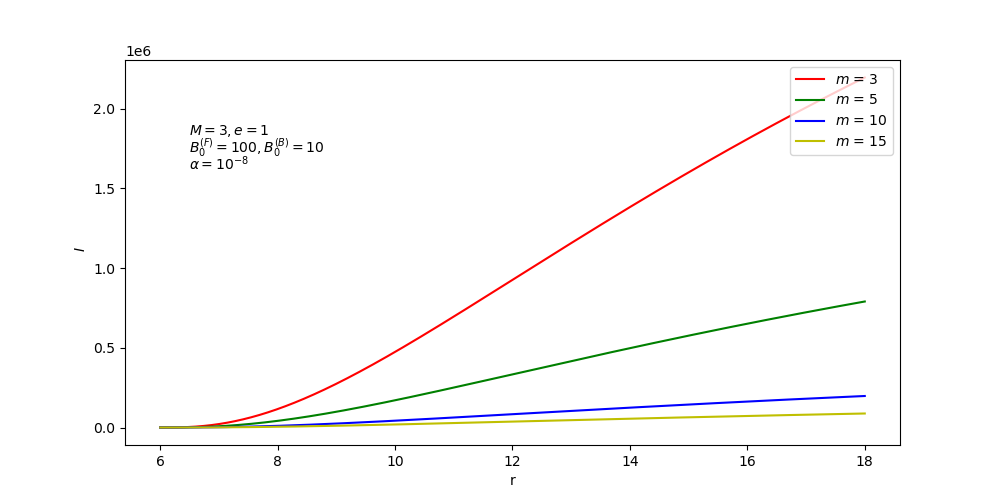} }
\caption{(color on line)  Dependence of intensity on the distance, for different values of a particle mass, for the dominant value of Maxwell field. }
\label{Fig:15}
\end{figure}

One should mention that,
in Figs. 1-6 and Fig. 8 we have been interested in qualitative behavior of the studied quantities due to the influence of {\it dark sector} ($\alpha$ coupling constant dependence),
while in Figs. 7 and 9, one implements the value of the coupling constant between {\it visible} and {\it dark sectors} closer to the recent experimental estimations \cite{mcc24}.

%%%%%%%%%%%%%%%%%%%%%%%%%%%
\section{Motion and energy loss}
This section will be devoted to the studies of circular motion in the equatorial plane of the charge particle emitting
synchrotron radiation. Using the Killing vector fields which admits the spacetime under scrutiny we define conserved quantities, then we
pay attention to the influence of {\it dark matter} sector on the energy loss of the studied particles.

\subsection{Motion of a charged particle}
For the motion in the spacetime under consideration we have two Killing vectors $\xi^\mu_{(t)}$ and $\xi^\mu_{(\phi)}$, which imply that two conserved quantities exist. They
are respectively, energy and azimuthal angular momentum. For elaborated case of equatorial motion they yield
\ben \label{def1}
\cE &=& - \frac{1}{m} \xi^\mu_{(t)} {\tilde p}_\mu =\frac{dt}{d \tau}~f(r),\\ \label{def2}
l_z &=& \frac{1}{m} \xi^\mu_{(\phi)}{\tilde p}_\mu = r^2 \Big( \frac{d \phi}{d \tau} + \te_A \tB_0^{(\tF)} + {\te_B} \tB_0^{(\tB)} \Big),
\een
where the dot is connected with the derivation with respect to time,  $d/ d\tau$. In the above relation we have denoted
\be
\tB_0^{(\tF)} = \frac{e}{2m} B_0^{(\tF)}, \qquad \tB_0^{(\tB)} = \frac{e_d}{2m} B_0^{(\tB)},
\label{stat1}
\ee
and
\be
B^{(\tF)}_0 = \frac{\sqrt{2 - \alpha}}{2} \Big( B^{(F)}_0 - B^{(B)}_0 \Big), \qquad
B^{(\tB)}_0 = \frac{\sqrt{2 + \alpha}}{2} \Big( B^{(F)}_0 + B^{(B)}_0 \Big).
\ee
Having in mind the normalization condition of the four-velocity, $u_\alpha u^\alpha = -1$, one arrives at
\be
\cE^2 = \Big(  \frac{d r}{d \tau} \Big)^2 + U_{eff},
\label{ee}
\ee
where the effective potential is provided by
\be
U_{eff} = f(r) \Big[ 1 + r^2 \Big( \frac{l_z}{r^2} - \te_A \tB_0^{(\tF)}  -  \te_B \tB_0^{(\tB)} \Big)^2 \Big].
\ee
For the case when $e_d =0$, it reduces to
\be
U_{eff} = f(r) \Bigg[ 1 + r^2 \Bigg( \frac{l_z}{r^2 } - {e}\Big( w^{(F)} + \frac{\alpha}{2} w^{(B)} \Big) \Bigg)^2
 \Bigg],
\ee
where one sets
\be
w^{(F)} = \frac{B_0^{(F)}}{2 m}, \qquad w^{(B)} = \frac{B_0^{(B)}}{2 m}.
\ee
The relation (\ref{ee}) is constraint, i.e., it is satisfied at the initial starting time and with the passage of time is always fulfilled, guiding the dynamics of 
$r(\tau)$ and $\theta( \tau)$.

The $r$ component of the equation of motion (\ref{eqmot}) is given by
\ben \nonumber
 \frac{d^2 r}{d \tau^2}
&+&  
+ \frac{l_z^2}{r^3}  \Big(  - f(r) + \frac{r f'(r)}{2} \Big)
+ \Big(  \te_A \tB_0^{(\tF)}  + \te_B  \tB_0^{(\tB)} \Big)^2 ~\Big(  f(r) + \frac{r f'(r)}{2} \Big) r  \\ 
&+& f'(r)
\Big[ \frac{1}{2} -  l_z \Big( \te_A \tB_0^{(\tF)}  +  \te_B \tB_0^{(\tB)} \Big) \Big] = 0, 
%{\ddot \theta} &=& - \frac{2}{r} ~{\dot r} {\dot \theta} + \frac{l_z^2 \cos \theta}{r^4 \sin^3 \theta} - \Big( \te_A \tB_0^{(\tF)}  +  \te_B \tB_0^{(\tB)} \Big)^2 \sin \theta \cos \theta.
\een
where $'$ denotes differentiation with respect to $r$-coordinate.

Restricting our consideration to the case $e_d=0$, one obtains the following equations:
\ben \nonumber
 \frac{d^2 r}{d \tau^2}
 &+& 
+ \frac{l_z^2}{r^3} \Big(  - f(r) + \frac{r f'(r)}{2} \Big)
+ {e^2}\Big( w^{(F)}  + \frac{\alpha}{2} w^{(B)} \Big)^2 ~\Big(  f(r) + \frac{r f'(r)}{2} \Big) r \\ 
&+& f'(r)
\Big[ \frac{1}{2} - {e~ l_z}\Big( w^{(F)}  +  \frac{\alpha}{2} w^{(B)} \Big) \Big] = 0, %\\
%{\ddot \theta} &=& - \frac{2}{r} ~{\dot r} {\dot \theta} + \frac{l_z^2 \cos \theta}{r^4 \sin^3 \theta} - 
%{e^2}
%\Big( w^{(F)}  +  \frac{\alpha}{2} w^{(B)} \Big)^2 \sin \theta \cos \theta.
\een
From the above it can be clearly seen that {\it dark sector} modifies the geodesic equations by its  own magnetic field and $\alpha$-coupling constant.

%%%%%%%%%%%%%%%%%%%%%%%%%%%%
\subsection{Energy loss due to synchrotron radiation}
Having in mind the relations (\ref{rad}) and (\ref{def1}), we can find the rate of energy loss for the particle moving around black hole, with respect to its proper time.
Namely, one obtains
\be
{\dot E} = - \beta_{Maxwell}~E^3 - \beta_{dark~ matter} ~E^3,
\label{loss}
\ee
where we have denoted
\ben
\beta_{Maxwell} &=& \frac{2 e^4}{3 m^5} B_0^{(F) 2}, \\
\beta_{dark ~matter} &=& \frac{2 e^4}{3 m^5}~\alpha~ B_0^{(B)} \Big( B_0^{(F)} + \frac{\alpha}{4} B_0^{(B)} \Big).
\een
The equation (\ref{loss}) shows that the particle energy loss increases with the proper time, due to emission of synchrotron radiation and {\it dark sector}
has its own share in it. The rate of energy loss depends on the factors $\beta_{Maxwell,~ dark ~matter}$. Namely for the established mass of the particle,
the increase of the values of magnetic fields, causes heavier loss of energy. On the other hand, for more massive particle the loss of energy is smaller.

%%%%%%%%%%%%%%%%%%%%%%%%%%%%%%%%%%%%%%%%%%%%%%%%%%  
%%%%%%%%%%%%%%%%%%%%%%%%%%%%%%%%%%%%%%%%%%%%%%%%%
\section{Conclusions}

In this paper we have considered synchrotron radiation of a massive charged particle moving in an equatorial plane around spherically symmetric weakly magnetized 
black hole. The particle is charged under two $U(1)$-groups, i.e., Maxwell and {\it dark photon} one. On the other hand, the black hole in question is also 
magnetized with Maxwell magnetic and {\it dark sector} fields. The magnetic fields were found by means of
Wald's procedure \cite{wal74}, taking into account {\it hidden sector} equations of motion and symmetries of the underlying
spacetime (Killing vectors).
In our considerations we implemented the model of {\it hidden sector} composed of two $U(1)$-gauge fields, Maxwell and {\it dark photon}, which interact between themselves by the 
so-called {\it kinetic mixing term}. 

The main aim of our studies is to trace the differences in synchrotron radiation studied in Einstein-Maxwell case \cite{sho15}, in comparison to the influence of {\it dark photon}.
Among all, it happens that the total radiation intensity of an ultra-relativistic particle is greater in comparison to the Maxwell case.
The intensity enlargement is proportional to $\alpha$ and $\alpha^2$-coupling constant between those two sectors. The more massive particle we consider, the lower the
intensity of radiation one can record. Moreover, intensity is proportional to the fourth power of the particle charge, as in ordinary Maxwell electrodynamics.

We were also interested in a qualitative  $\alpha$-coupling influence on the studied quantities, like intensity of radiation and $\omega_c$, taking into account
various magnitude of ordinary and {\it dark photon} magnetic fields. One can conclude that the larger value of {\it dark matter} magnetic field
we consider, the bigger distances among curves responsible for various $\alpha$-values is observed.
All the studied quantities growth with constant coupling growth. On the other hand,
when one considers the mass of the moving particle, we can conclude that the smaller the mass we have, the larger the value of intensity we achieve (for fixed black hole mass, particle charge and $\alpha$-value).

As far as the loss of energy is concerned, it depends on the auxiliary term proportional $\alpha$-coupling constant and it is a function of Maxwell and {\it dark photon}
magnetic fields. For massive particles the loss of energy caused by emission of synchrotron radiation is smaller, comparing to the depletion of energy for light ones.

Having in mind that {\it dark photon} field intensifies the magnitude of synchrotron radiation, one can look for such phenomena in the nearby of black holes.

To apprehend the physical significance of this problem, one can consider the model of  magnetar/neutron star moving around a supermassive black hole in the centre of a galaxy. In order to weakly magnetize the black hole, we consider the orbit of revolution to be significantly larger than the radius of the supermassive black hole. On the other hand, one can safely infer magnetar/neutron star
 to be a point particle in comparison to the supermassive black hole. This point particle in a curved path would lose energy via electromagnetic radiation which is dominated by time-varying dipole moment as well as gravitational radiation dominated by the time-varying quadrupole moment. As a result of energy loss, the orbit of the magnetar/neutron star would shrink and it would eventually merge into the supermassive 
black hole. 

On the other hand, both theoretical \cite{hei97} and observational evidences \cite{m87}
reveal that supermassive black hole in M 87 galaxy gives rise to synchrotron radiation in jets generated by the gravitational acceleration of ions in their polar magnetic fields.
Observations conducted by Hubble Space Telescope confirmed the predictions showing a bluish tint in the jet in question.
Now M87 galaxy is a laboratory for testing black hole astrophysics and observational knowledge across various wavelength give us opportunity to examine physics of jets and central objects with unprecedented details. The accumulated data may shed some light on further {\it dark matter} searches.

Apart from using the gravitational lensing technique applicable in 'seeing' the distribution of {\it dark matter}, 
%we obtain the additional tool for conducting the future experiments by which we may be able to search for
%%%%%
maybe in the near future more detailed analysis of synchrotron radiation in the vicinity of black holes and
future sophisticated experiments conducted by Event Horizon Telescope collaboration, will enable us
to find some differences in the aforementioned radiation characteristics, comparing to ordinary visible sector electrodynamics. They may constitute 
the first trace in search for
%%%%%
this illusive ingredient of our Universe.

%%%%%%%%%%%%%%%%%%%%%%%%%%%%%%%%%%%%%%%%%%%%%%%%%%%%%%%%%%%%%%%%%%
%%%%%%%%%%%%%%%%%%%%%%%%%%%%%%%%%%%%%%%%%%%%%%%%%%%%%%%%%%%%%%%%%%
%%%%%%%%%%%%%%%%%%%%%%%%%%%%%%%%%%%%%%%%%%%%%%%%%%%%%%%%%%%%%%%%%%%%%%%%%%%%%%
\acknowledgments
M. R. and P. V. were partially supported by Grant No. 2022/45/B/ST2/00013 of the National Science Center, Poland.

%%%%%%%%%%%%%%%%%%%%%%%%%%%%%%%%%%%%%%%%%%%%%%%%%%%%%%%%%%%%%%%%%%%%%%%%%%%%%%

%%%%%%%%%%%%%%%%%%%%%%%%%%%%%%%%%%%%%%%%%%%%%%%%%%%%%%%%%%%%%%%%%%

%%%%%%%%%%%%%%%%%%%%%%%%%%%%%%%%%%%%%%%%%%%%%%%%%%%%%%%%%%%%%%%%%%%%%%%%%%%%%%%

\end{document}